\documentclass[fleqn,usenatbib]{mnras}

\usepackage{newtxtext,newtxmath}

\usepackage[T1]{fontenc}

\DeclareRobustCommand{\VAN}[3]{#2}
\let\VANthebibliography\thebibliography
\def\thebibliography{\DeclareRobustCommand{\VAN}[3]{##3}\VANthebibliography}

\usepackage{graphicx}	

\newcommand{\HMHD}{FID-MHD}

\newcommand{\HHD}{FID-HD}

\title[
Non-Kolmogorov ICM turbulence]{
Non-Kolmogorov turbulence in multiphase intracluster medium driven by cold gas precipitation and AGN jets}

\author[Ch. Wang et al.]{
C. Wang,$^{1}$\thanks{E-mail: wangcha@umich.edu (CW)}
M. Ruszkowski,$^{1}$
C. Pfrommer,$^{2}$
S. Peng Oh$^{3}$
and H.-Y. K. Yang$^{4}$
\\
$^{1}$Department of Astronomy, University of Michigan, 1085 S. University Avenue, 311 West Hall, Ann Arbor, MI 48109, USA\\
$^{2}$Leibniz-Institut f\"ur Astrophysik Potsdam (AIP)
An der Sternwarte 16, 14482 Potsdam, Germany\\
$^{3}$Department of Physics, University of California, Santa Barbara, CA 93106, USA\\
$^{4}$Institute of Astronomy and Department of Physics, National Tsing Hua University, No. 101, Section 2, Kuang-Fu Road, Hsinchu 30013, Taiwan
}

\date{Accepted XXX. Received YYY; in original form ZZZ}

\newcommand{\uu}[1]{~{\rm #1}}
\newcommand{\lrmhd}{LR-MHD~}
\newcommand{\lrhd}{LR-HD~}
\newcommand{\hrhd}{FID-HD~}
\newcommand{\hrmhd}{FID-MHD~}

\pubyear{2020}

\begin{document}
\label{firstpage}
\pagerange{\pageref{firstpage}--\pageref{lastpage}}
\maketitle

\begin{abstract}
Active galactic nuclei (AGN) feedback is responsible for maintaining plasma in global thermal balance in extended halos of elliptical galaxies and galaxy clusters. Local thermal instability in the hot gas leads to the formation of precipitating cold gas clouds that feed the central supermassive black holes, thus heating the hot gas and maintaining global thermal equilibrium.
We perform three dimensional magnetohydrodynamical (MHD) simulations of self-regulated AGN feedback in a Perseus-like galaxy cluster with the aim of understanding the impact of the feedback physics on the turbulence properties of the hot and cold phases of the intracluster medium (ICM).
We find that, in general, the cold phase velocity structure function (VSF) is steeper than the prediction from Kolmogorov's theory. We attribute the physical origin of the steeper slope of the cold phase VSF to the driving of turbulent motions primarily by the gravitational acceleration acting on the ballistic clouds.
We demonstrate that, in the pure hydrodynamical case, the precipitating cold filaments may be the dominant agent driving turbulence in the hot ICM. The arguments in favor of this hypothesis are that: (i) the cold phase mass dominates over hot gas mass in the inner cool core; (ii) hot and cold gas velocities are spatially correlated;
(iii) both the cold and hot phase velocity distributions are radially biased. 
We show that, in the MHD case, the turbulence in the ambient hot medium (excluding the jet cone regions) can also be driven by the AGN jets. The driving is then facilitated by enhanced coupling due to magnetic fields of the ambient gas and the AGN jets. In the MHD case, turbulence may thus be driven by a combination of AGN jet stirring and filament motions.
We conclude that future observations, including those from high spatial and spectral resolution X-ray missions, may help to constrain self-regulated AGN feedback by quantifying the multi-temperature VSF in the ICM.

\end{abstract}

\begin{keywords}
intracluster medium  -- AGN feedback -- turbulence -- MHD
\end{keywords}

\section{Introduction}
\begin{figure*}
  \begin{center}
    \leavevmode
    \includegraphics[width=0.9\textwidth]{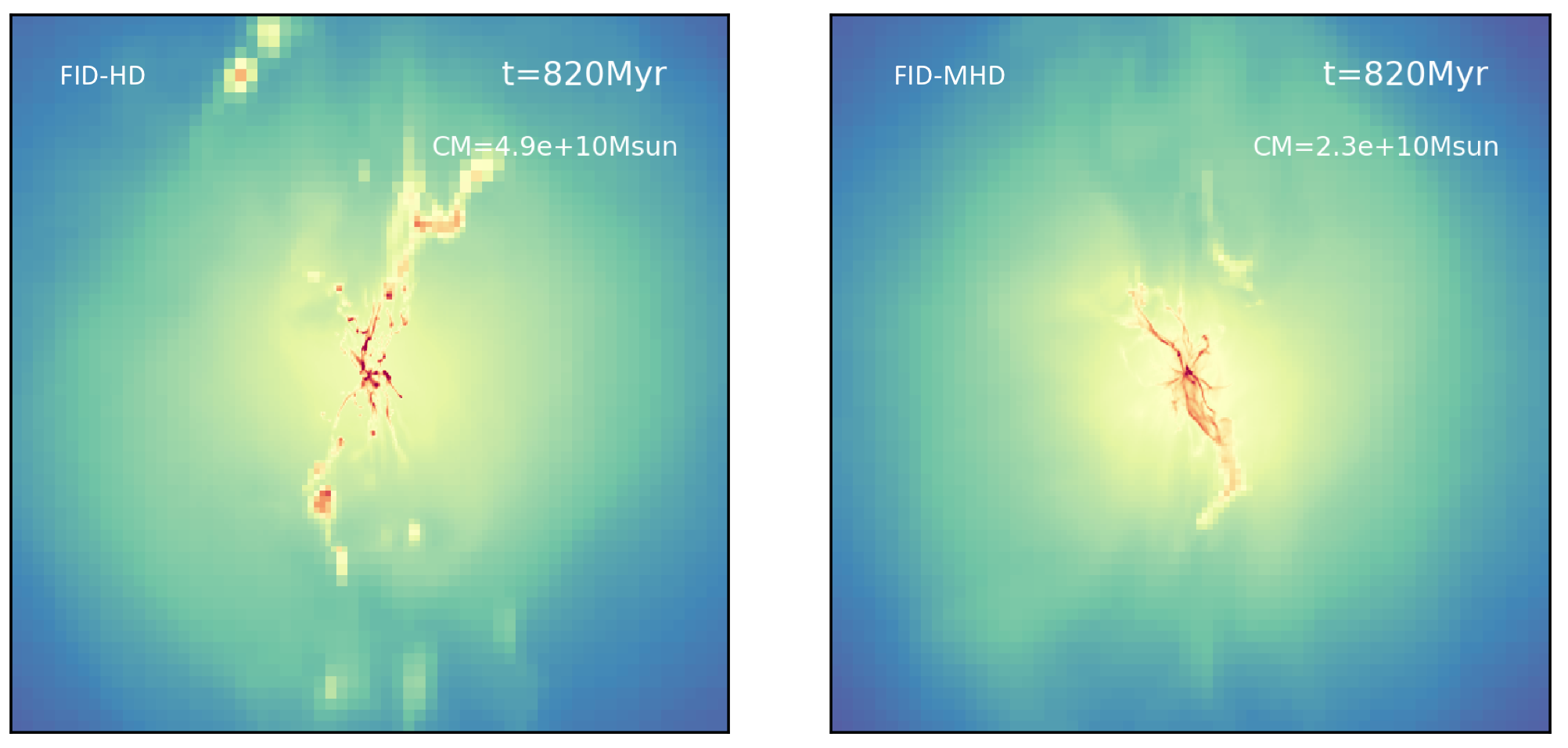}  
       \caption[]{Snapshots of the projected gas density from our fiducial hydrodynamical (\HHD; left) and fiducial MHD (\HMHD; right) runs. The projection is done along the $x$-axis and within the central 125\uu{kpc}-wide cube. It is clear that cold filaments are more spatially confined in \HMHD than \HHD.   
       }
     \label{fig:projDensity}
  \end{center}
\end{figure*}

Hot and dilute plasmas in elliptical galaxies and galaxy clusters often exhibit cooling times shorter than the Hubble time. These short cooling times lead to local thermal instability in the hot atmospheres of these objects. The observational evidence for this instability comes in the form of detection of molecular and H$\alpha$+[N II] emission from dense and cold gas 
\citep[e.g.,][]{Lakhchaura2018, Pulido2018, Babyk2019, Olivares2019}. The systems containing the cold gas are characterized by relatively shorter radiative cooling times, higher overall gas density, lower entropies, and less symmetric X-ray emissivity distributions caused by gas motions \citep[e.g.,][]{Lakhchaura2018}. While there continues to be a debate on whether in the locally thermally unstable systems the ratios of the cooling time to the free-fall time are systematically lower (see \citet{Lakhchaura2018} for evidence in favor of this hypothesis and \citet{Babyk2019} for the opposing view), there is emerging consensus that the cold gas has cooled out of the hot plasma.\\
\indent
Since the cool and thermally unstable gas is expected to be unsupported by the pressure gradients in the hot gas phase, the cold phase precipitates and feeds the central supermassive black holes \citep[e.g.,][]{Werner2013,Werner2014}. There is well-established evidence for active galactic nuclei feedback from black holes in the systems characterized by short radiative cooling times. For example, recent LOFAR observations by \citet{Birzan2020} of X-ray cavity systems present in such systems confirm that there is a connection between radio emission and AGN jet power inflating the cavities.\\
\indent
Observations of cold gas precipitation and AGN feedback in these systems have spurred theoretical investigations aiming to explain the feedback cycle. Using idealized numerical experiments in which heating was distributed globally to balance cooling in the time average sense, \citet{McCourt2012, Sharma2012, GRO13} demonstrated that heated atmospheres can remain in global thermal equilibrium while exhibiting local thermal instability. In these models, thermal instability led to precipitation of cold gas, feeding of the central engine, and subsequent heating to balance radiative cooling globally. Using analytic methods to interpret recent simulations of feedback, \citet{Voit2017} considered two condensation modes -- (i) precipitation and infall and (ii) uplift of ambient gas by outflows followed by condensation -- and argued in favor of the former. An alternative suggestion was made by \citet{McNamara2016} who proposed that 
cold clouds condense out of the low-entropy gas uplifted by the AGN bubbles. \\ 
\indent
Hydrodynamical AGN feedback simulations including bipolar jet outflows were performed by 
\citet{GRS2012, Li2017, Martizzi2019, Wang2019}. These simulations were successful in establishing (i) self-regulating feedback cycle, where the balance of cooling and heating was achieved globally, and where the amount of energy injected by the jets was regulated by the amount of accretion of the cold gas, and (ii) predicting density, temperature, and entropy profiles in agreement with the observations. In particular, the simulations by \citet{Wang2019} of self-regulated AGN feedback can maintain the observed properties of single and multi phase halos, which is consistent with analytical models \citep{Voit2015} and recent observations \citep{Frisbie2020}. Using a similar approach, albeit excluding the formation of the cold phase clouds, \citet{Yang2016} performed a careful analysis of various contributions to the AGN heating budget. Specifically, they demonstrated that a substantial contribution to heating comes from shock heating and turbulent mixing (inside the jet cones) and weak shocks and adiabatic compression (outside the jet cones).
The importance of sound wave heating was recently put on a firmer footing (\citealt{Bambic2019}, see also \citealt{RBB2004a,RBB2004b}).\\
\indent
The nature of the simulated AGN feedback cycle and the properties of the multiphase gas depend on the physics included in the simulations. Recent results by \citet{Beckmann2019}
demonstrate that while both the precipitation and uplift of dense gas is present in the simulated atmospheres, purely hydrodynamical simulations struggle to regulate the cluster cooling-feedback cycle and lead to very clumpy distributions on cold gas that is inconsistent with the very filamentary cold structures seen in the observational data. These findings underscore the importance of investigating the impact of a wider array of physical processes in the simulations. \citet{Qiu2019} performed hydrodynamical simulations of self-regulated feedback including the effects of radiation feedback, and while they observed the formation of elongated filaments, the simulations also resulted in the formation of overly massive cold central disks. \\
\indent
MHD simulations of AGN jet feedback with super-Lagrangian resolution and including the effect of cosmic ray (CR) pressure were performed by \citet{Weinberger2017}. Single injection AGN events were studied using CR MHD simulations including CR diffusion and Alfv\'en wave cooling, thus emulating CR streaming, by \citet{Ehlert2018}, who concluded that CR heating rates were significant compared to cooling and matched radial CR pressure profiles of one-dimensional steady-state CR heating models (\citealt{Jacob2017a,Jacob2017b}, see also \citealt{Guo2008}). The CR pressure of AGN lobes is also consistent with the missing thermal pressure as inferred from Sunyaev-Zel'dovich observations of the extreme AGN outburst in MS0735 \citep{Abdulla2019,Yang2019,Ehlert2019}. CR MHD simulations of self-regulated AGN jet feedback cycle including the effects of CR streaming, and associated with it CR heating of the gas, were performed by \cite{RYR17} and \cite{Wang2020}, who demonstrated that CR could serve as a dominant heating agent to keep the atmospheres in global thermodynamical equilibrium. \cite{Wang2020} also demonstrated that the magnetic fields play a crucial role in angular momentum redistribution via magnetic breaking of the precipitating cold gas, which allows the gas to accrete and feed the central supermassive black hole without invariably forming massive cold central tori. More broadly, this suggest that the magnetic fields may play a role in shaping the statistical properties of the velocity distribution of the gas in these systems. \\
\indent
As mentioned above, AGN feedback, cold gas precipitation, and the physics relevant to the problem of the feedback cycle are all intricately related to the question of turbulence in the atmospheres of galaxies and clusters. Thus turbulence measurements could provide constraints on how the AGN feedback works in realistic systems. 
A recent review of turbulence in the hot halos of ellipticals and clusters has been presented by \citet{Simionescu2019}.
Constraints on the level of turbulence have been obtained from non-spatially-resolved line broadening. \citet{Sanders2013} reported velocity limits of 300 to 500 km s$^{-1}$ in elliptical galaxy atmospheres. Using a combination of resonant scattering and line broadening, \citet{Ogorzalek2017} measured turbulent velocities in the hot halos of giant ellipticals and found typical Mach numbers of $\sim 0.45$. Direct measurements of the level of turbulence in the Perseus cluster obtained using the Hitomi mission \citep{Hitomi2016, Hitomi2018} are consistent with relatively low level
of turbulence in this cluster. Future high spectral and spatial resolution X-ray missions may be able to quantify the level of turbulence in significantly more detail by measuring the velocity power spectrum or VSF of the hot gas phase \citep{Zhuravleva12,ZuHone16}. Recently, a very promising alternative approach to constraining turbulence has been presented by \citet{Li20}, who measured the VSF of the cold gas phase and showed that the slope of the cold phase VSF departs from the Kolmogorov prediction. Motivated by these observations, \citet{Hillel2020} simulated the VSF in non-radiative simulation of AGN jet outburst
and concluded that the VSF is steeper than that expected based on the Kolmogorov theory of turbulent cascade. 
In this paper, we study via MHD simulations of the self-regulated AGN feedback, the properties of turbulence in the hot and cold phases of the intracluster medium. In particular, we discuss the coupling between the two gaseous phases and the mechanisms driving turbulence in these phases, and make predictions for the slope of the cold phase VSF, and its relationship to the hot phase counterpart. The paper is organized as follows. In Section 2 we present the simulation approach. The discussion of the results is presented in Section 3, and we conclude in Section 4.  

\section{Methods}

\begin{table}
\caption{List of models}
\label{table:listOfModels}
\begin{tabular}{ccccccccc} 
\hline
 name            & magnetic field & $n_{\rm max}$ &  $\Delta x_{\rm min}$ \\
 \hline 
 \lrhd           & no             &  6         &  $0.244\uu{kpc}$  \\
 \lrmhd          & yes            &  6         &  $0.244\uu{kpc}$   \\
 \hrhd           & no             &  7         &  $0.122\uu{kpc}$   \\
 \hrmhd          & yes            &  7         &  $0.122\uu{kpc}$   \\
 \hline
 \end{tabular}
\end{table}

\begin{figure}
  \begin{center}
    \leavevmode
    \includegraphics[width=\columnwidth]{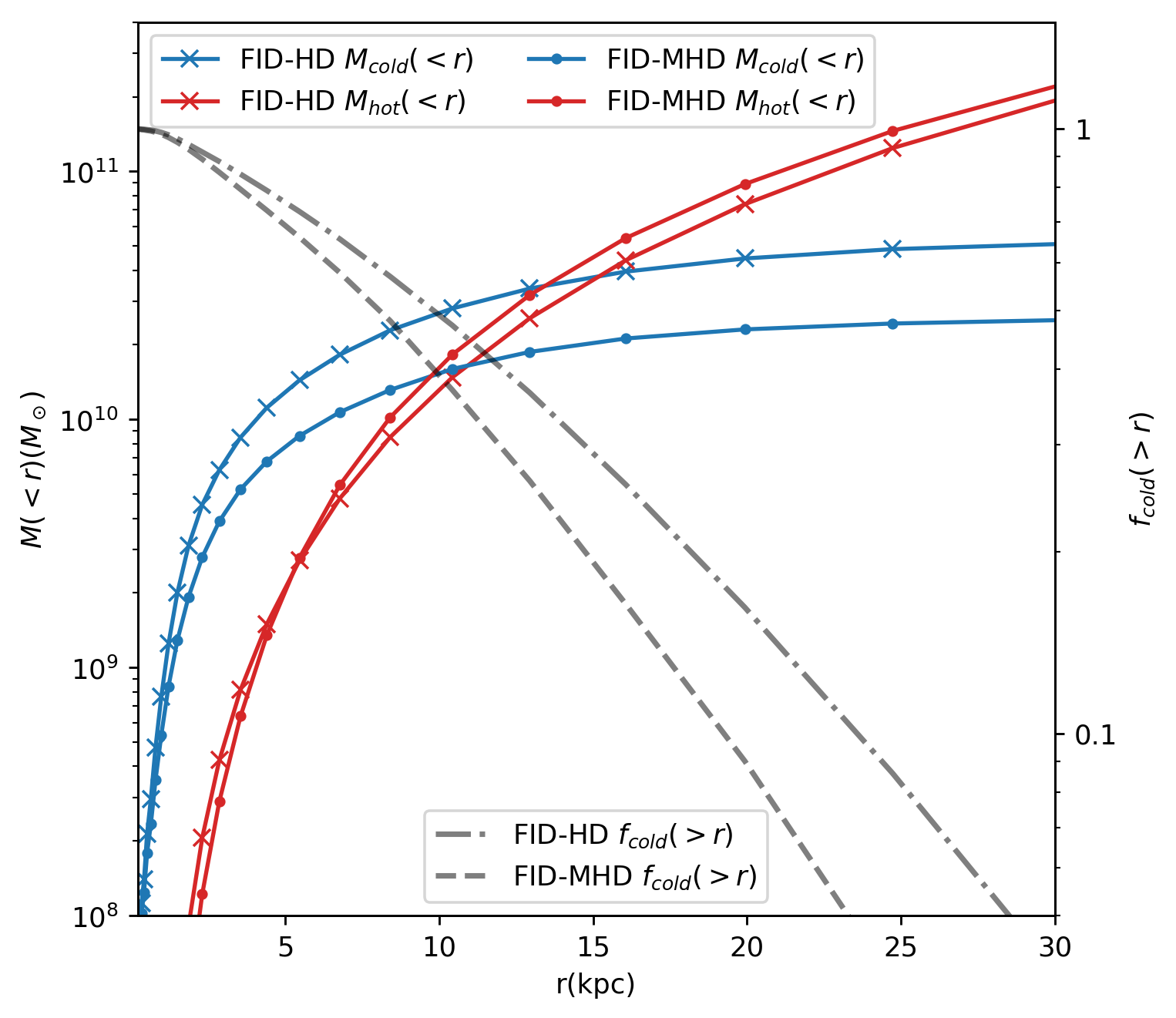}  
       \caption[]{ Blue and red lines: radial profile of enclosed mass ($M<(r)$) of hot and cold gas in \HHD\ and \HMHD\ cases.
       Grey lines: the profiles of cold gas fraction outside of a given radius ($f_{\rm cold}(>r)$ ). All profiles are averaged over $t=0.4\sim1.5$ Gyr. }
     \label{fig:massDist}
  \end{center}
\end{figure}

\begin{figure*}
  \begin{center}
    \leavevmode
    \includegraphics[width=0.95\textwidth]{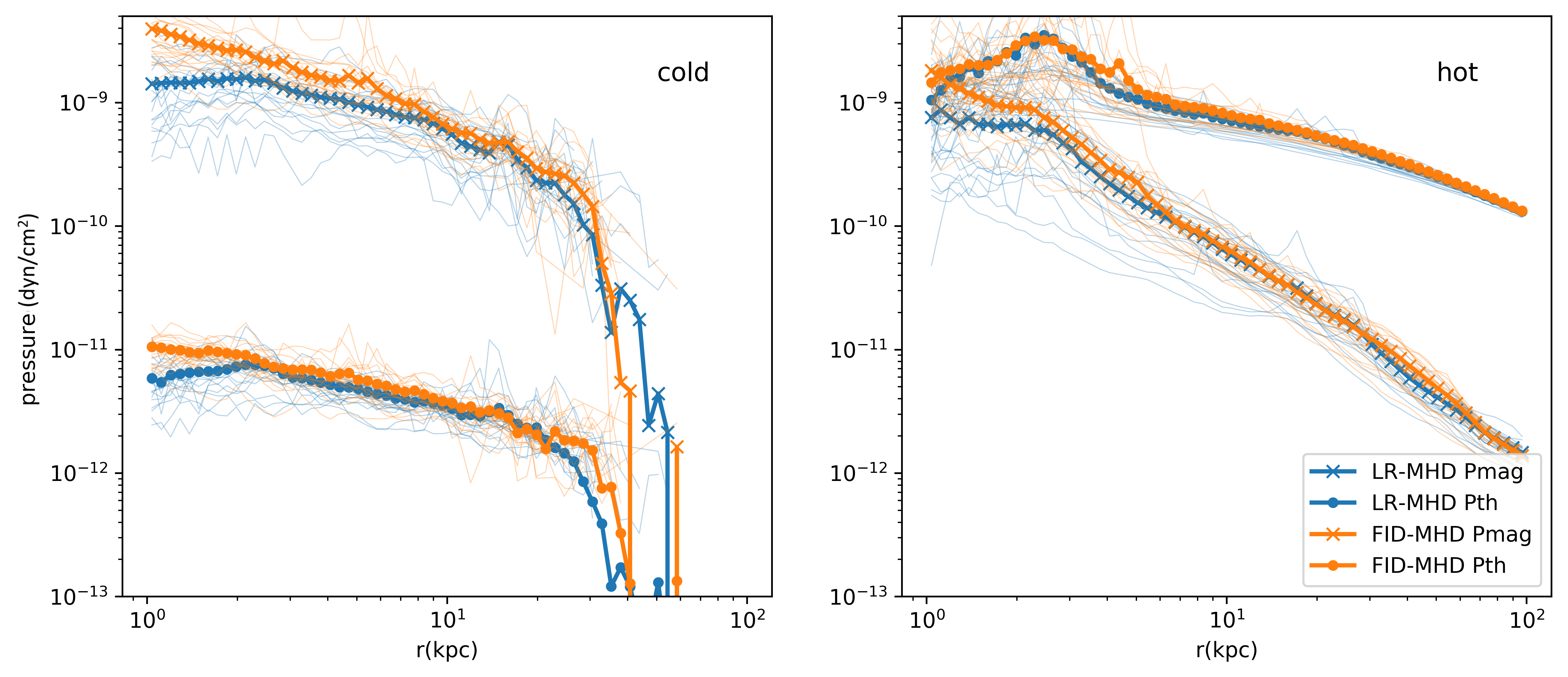}  
       \caption[]{Left: radial profile of the magnetic pressure ("x"-labelled line) and thermal pressure (dot-labelled line) of the cold gas in the MHD runs.
Right: same as on the left but for the hot gas.}
     \label{fig:pressures}
  \end{center}
\end{figure*}

We perform three-dimensional MHD simulations using the FLASH code \citep{2000ApJS..131..273F, dubey2008introduction} using the directional unsplit staggered mesh MHD solver \citep{Lee09, Lee13}. 
We adopt a simulation domain with statically-refined grids. Similar to \cite{Wang2020}, the domain is refined by a set of nested cubic regions. The entire domain is a cubic region with width $L_{\rm box}=1\uu{Mpc}$ and $64^3$ base grids. The central regions with width 
\begin{equation}\label{eq:nestsWidth}
L_n=L_{\rm box}/2^{n}\uu{kpc} 
\end{equation}
have $n$ additional nested refinement levels. We include models with different $n_{\rm max}$, as listed in Table \ref{table:listOfModels}. The size of the smallest cell is $\Delta x_{\rm min}=L_{\rm box}/64/2^{n_{\rm max}}$. For the resolution study, we modify the width of the two most refined regions in the high resolution models, so that
$L_7$ in the high resolution models equals $L_6\approx15.6\uu{kpc}$ in the low resolution models.
Therefore, in order to calculate the sizes of the regions at a given refinement level 
in the \hrhd and \hrmhd cases,
we use eq.~\eqref{eq:nestsWidth} for $n\le 5$ and set $L_7\approx15.6\uu{kpc}$ and choose $L_6$ such that it falls between $L_5$ and $L_7$ (our adopted value of $L_6$ is $\approx23.4\uu{kpc}$). \\
\indent
We use the diode boundary conditions, where all variables have zero gradient and gas can only flow out at the domain boundary.

\subsection{Cluster initial conditions}
We setup the initial conditions of the hot gaseous halo consistent with the Perseus cluster. 
For the temperature profile, we use an analytical fit based on the observed X-ray surface brightness distribution in the Perseus cluster \citep{2003ApJ...590..225C}:
\begin{equation}\label{eq:tempProfile}
T(r) = 7\uu{keV} \frac{1+(r_{\rm kpc}/71)^{3}}{2.3+(r_{\rm kpc}/71)^3} [1+(r_{\rm kpc}/380)^2]^{-0.23}.
\end{equation}
We include a static gravitational field with contributions from a dark matter halo and stars. The dark matter  potential is described by an NFW profile \citep{1996ApJ...462..563N}, with scale radius $r_s=358.3\uu{kpc}$, virial radius $r_{\rm vir}=2.44\uu{Mpc}$ and virial mass $M_{\rm vir}=8.5\times10^{14}M_{\odot}$. 
The gravitational acceleration due to stars is based on the analytic fit to the de Vaucouleurs profile \citep{2006ApJ...638..659M}:
\begin{equation}\label{eq:stellarG}
g_{\rm star}(r) = \left[\frac{r_{\rm kpc}^{0.5378}}{2.853\times10^{-7}} + \frac{r_{\rm kpc}^{1.738}} {1.749\times10^{-6}} \right]^{-1.11}\uu{cm~s^{-2}}.
\end{equation}
Assuming hydrostatic equilibrium, we then calculate the density profile and normalize it to match the azimuthally averaged observed density profile \citep{2006ApJ...638..659M}. 

Following \cite{Ruszkowski2007}, we set up tangled magnetic fields with power spectrum $B_k\propto k^{-11/6}{\rm exp}(-k^4/k_0^4)$, where $k_0=100(2\pi/L_{\rm box})$. To obtain magnetic fields with this power spectrum and plasma $\beta\sim100$, we first inversely Fourier transform the power spectrum to real space; then normalize the real space magnetic fields to have $\beta\sim100$; Fourier transform the magnetic fields; clean the magnetic field divergence in Fourier space; and finally perform inverse Fourier transformation to obtain real space magnetic fields. We repeat this procedure until the magnetic fields become divergence free. 
For radiative cooling, we adopt the tabulated Sutherland-Dopita cooling function assuming one-third solar metallicity \citep{SutherlandDopita1993}. 
\subsection{AGN feedback}
We use the cold accretion model to simulate the fueling of the AGN. The cold gas is accreted at a rate of $\dot{M}_{\rm acc} = M_{\rm acc}/5\uu{Myr}$, where $M_{\rm acc}$ is the total mass of the cold gas ($T<10^{5}\uu{K}$) within the depletion region, $r<1.2\uu{kpc}$. For each computational time step ($\Delta t$), $\dot{M}_{\rm acc}\Delta t$ of the cold gas is removed in the depletion region and loaded to the jet base. The gas in the jet base is then launched into the halo via bipolar jets along $z-$axis of the simulation domain. 
The jet base is a cylinder at the domain center with a radius of $1.25\uu{kpc}$ and a height of $4\uu{kpc}$. We assume a jet precession angle of 15 degrees and a period of $10\uu{Myr}$.  
The AGN feedback is purely kinetic with power $\dot{E}_k=\epsilon\eta \dot{M}_{\rm acc}c^2$, where $\epsilon=10^{-3}$ is the feedback efficiency, $\eta=1$ is the mass loading factor, and $c$ is the speed of light. 

\begin{figure*}
  \begin{center}
    \leavevmode
    \includegraphics[width=\textwidth]{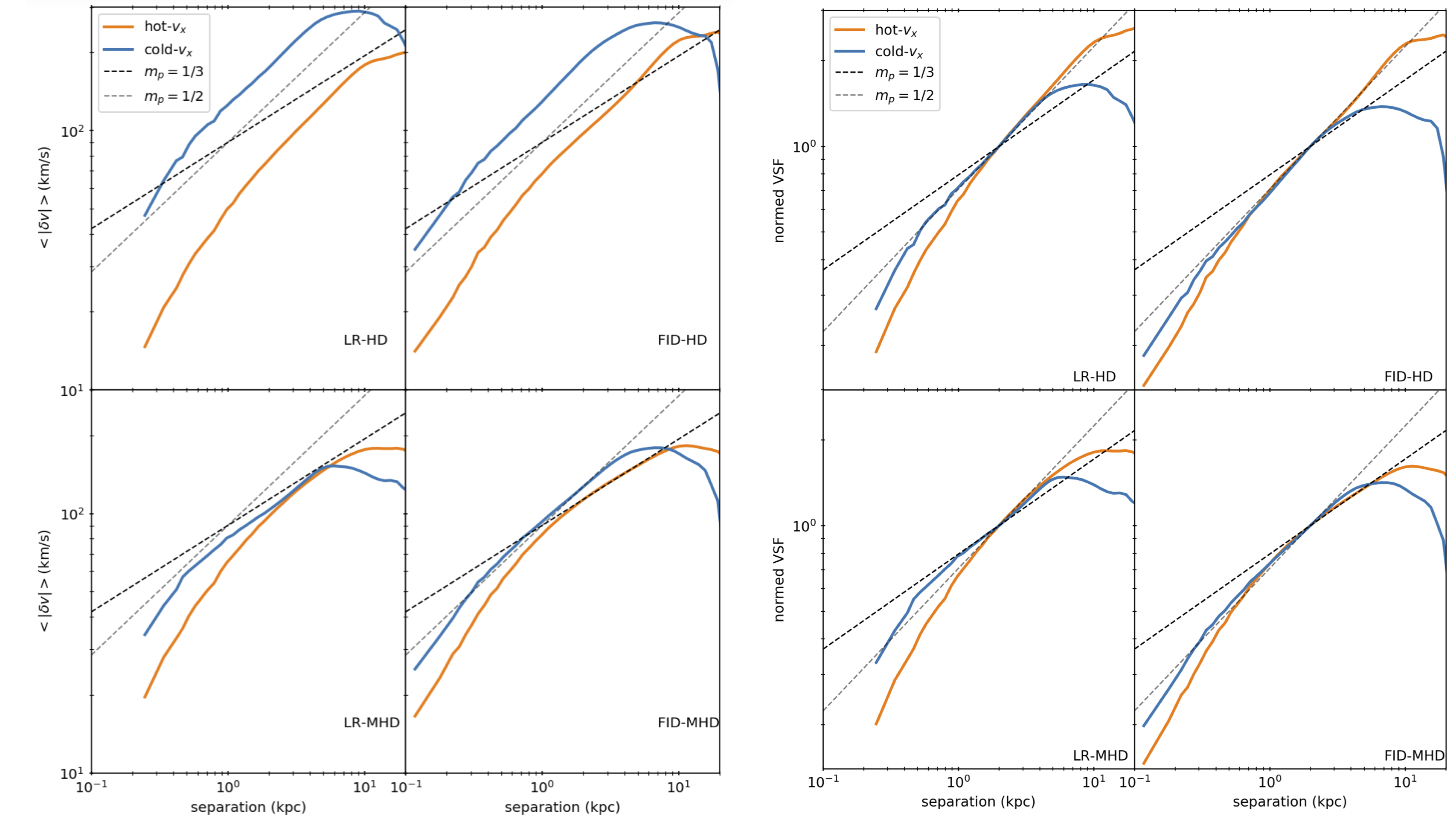}  
       \caption[]{VSFs of hot (orange) and cold (blue) phase gas averaged over $t=0.5\sim 1 \;{\rm Gyr}$  of the four runs. The right four panels are the same as the left four except that they are normalized to 1 at the separation of 2\uu{kpc}. The black and grey dashed lines have slopes $m_p=m_{\rm kol}=1/3$ and $m_p=1/2$, respectively. }
     \label{fig:vsfSepNames}
  \end{center}
\end{figure*}

\begin{figure}
  \begin{center}
    \leavevmode
    \includegraphics[width=\columnwidth]{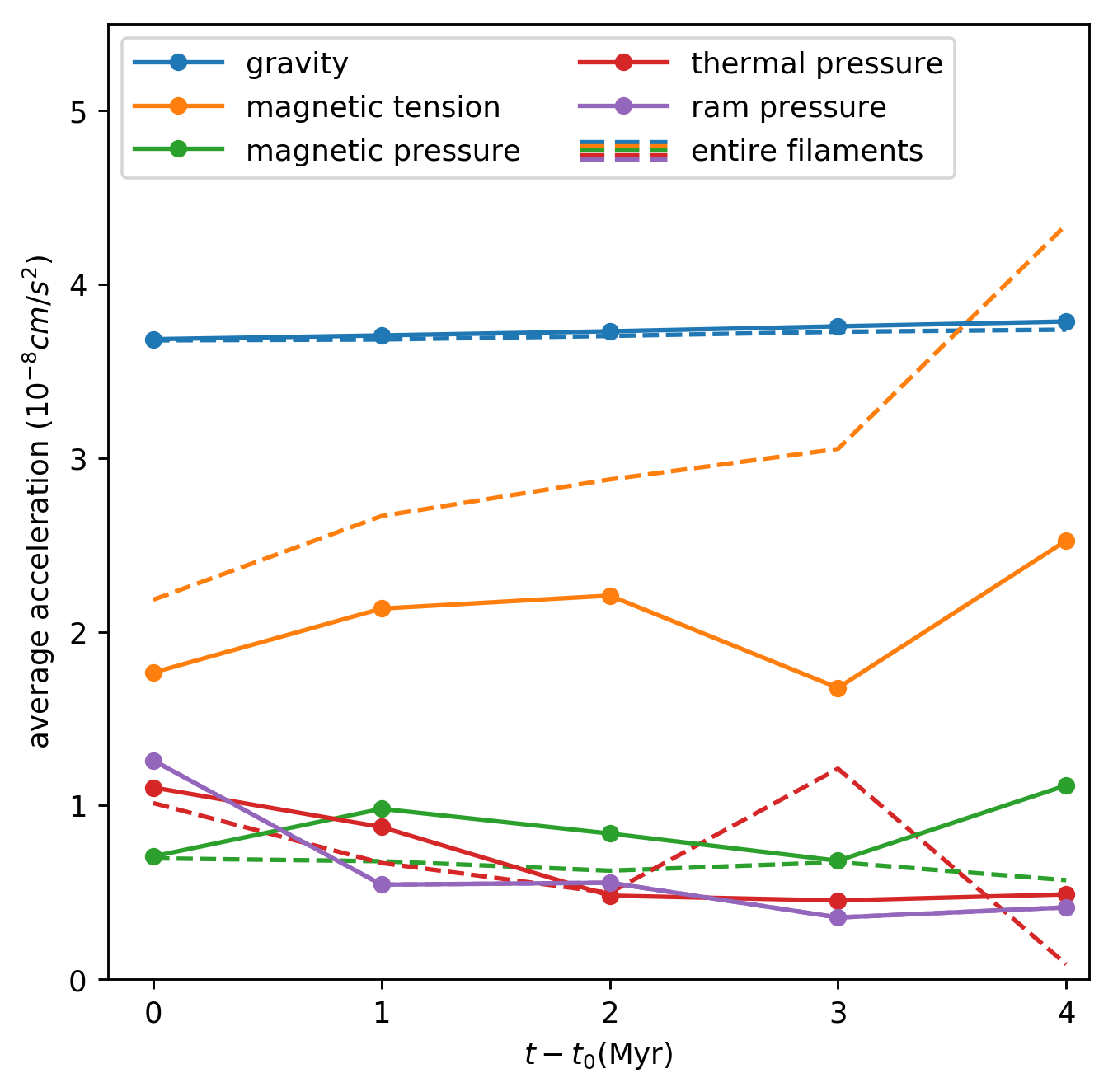}  
       \caption[]{Values of accelerations contributed by different sources during the infall process of a cold filament. The solid lines show the values averaged over the head of the filament and the dashed lines are over the entire filament.}
     \label{fig:eulerAnalysis}
  \end{center}
\end{figure}

\section{Results}
\subsection{General characteristics of cold gas distribution and the evolution of the cluster}

In all cases, the systems go through the self-regulated AGN feedback cycles. The development of thermal instabilities leads to cold gas condensation in the form of filaments. 
The cold gas forms a long-lived massive rotating disk during the long term evolution of the hydro cases, while the disk is absent in the MHD cases. The same phenomenon was found in \citet{Wang2020}. As we argued there, the formation of such a disk is unphysical and in tension with observations. The magnetic tension force can effectively decelerate the cold gas. Therefore the magnetic fields redistribute the angular momentum of the cold gas, and prevents disk formation.

Snapshots of the projected gas density are shown in Fig. \ref{fig:projDensity} in the \hrhd and \hrmhd case (left and right panel, respectively). There are clear morphological differences between these two cases. While in the hydro case, the dense and cold gas exists in the form of blobs and filaments, in the MHD case the dense gas structures are noticeably more filamentary. In both cases the filaments tend to be biased in the radial direction. The cold filaments are more spatially confined to the cluster center in the MHD cases than those in hydro cases (Fig. \ref{fig:projDensity}). 

In the \HMHD\ runs 95\% of the cold gas mass is contained within $r\la25$ kpc, while for \HHD, this happens within $r\la30$ kpc (grey lines in Fig. \ref{fig:massDist};  cold gas is defined as having $T<2\times10^{4}{\rm K}$ and $\rho >10^{-24}$g cm$^{-3}$, and hot gas as having $10^{7}{\rm K}<T<10^{8}{\rm K}$). When the system is active, the typical mass of the cold gas is $10^{10}\sim5\times10^{10}M_\odot$, which is broadly consistent with the high end of the molecular gas mass observed in galaxy clusters \citep[e.g.,][]{Vantyghem2018, Olivares2019}. This is especially so given that our definition of the cold gas includes the molecular gas as well as the gas at higher temperatures. We also note that, as evidenced in Fig. \ref{fig:massDist}, the time averaged amount of cold gas is smaller in the \hrmhd cases compared to the \hrhd case.

\indent
In the left panel of Fig. \ref{fig:pressures} we show radial profiles of the magnetic pressure ("x"-labelled line) and thermal pressure (dot-labelled line) of the cold gas in the MHD runs. The right panel shows the profiles of the same quantities but for the hot gas. It is clear from this figure that the pressure support of the cold filaments is dominated by the magnetic pressure. The magnetic pressure of the cold filaments is in balance with the thermal pressure of the ambient hot gas, which is evident from comparing top orange line in the left panel (representing magnetic pressure) with the top orange line in the right line (representing thermal pressure) -- both of these curves have very similar values in the central $\sim$20 kpc. This result is in agreement with the optical emission line observations of Perseus cluster \citep{Fabian2008} and numerical simulations \citep{Sharma2010}. Furthermore, the low plasma $\beta$ filaments are consistent with the models suggesting that the H$\alpha$ emission of the cold filaments is powered by magnetic reconnection \citep{Churazov2013} or cosmic rays \citep{Ruszkowski2018}. 

\subsection{Velocity structure function}
Throughout this paper, we use the first order VSF of the $x-$component to study the properties of the ICM turbulence. It is defined as VSF($l$)$\equiv \langle |{v_x}({\mathbf r}+\mathbf{l})-v_x(\mathbf{r})|\rangle$, where $v_x$, $\mathbfit{r}$ and $\mathbfit{l}$ are the $x-$component of the gas velocity, position vector and the vector connecting a pair of points, respectively, and the averaging is performed over pairs of points with the same separation $l=|\mathbfit{l}|$. We note that the $y$-component VSF shows qualitatively the same results due to the axisymmetry about the jet launching $z-$axis. We denote the slope of the VSF as $m_p$, i.e., ${\rm VSF}(l)\propto l^{m_p}$. In this paper, we focus on three-dimensional VSF and on how the physics included in the simulations affects the VSF slope. Thus, our results can be compared to the prediction from the Kolmogorov theory in a straightforward fashion. 

\begin{figure}
  \begin{center}
    \leavevmode
    \includegraphics[width=\columnwidth]{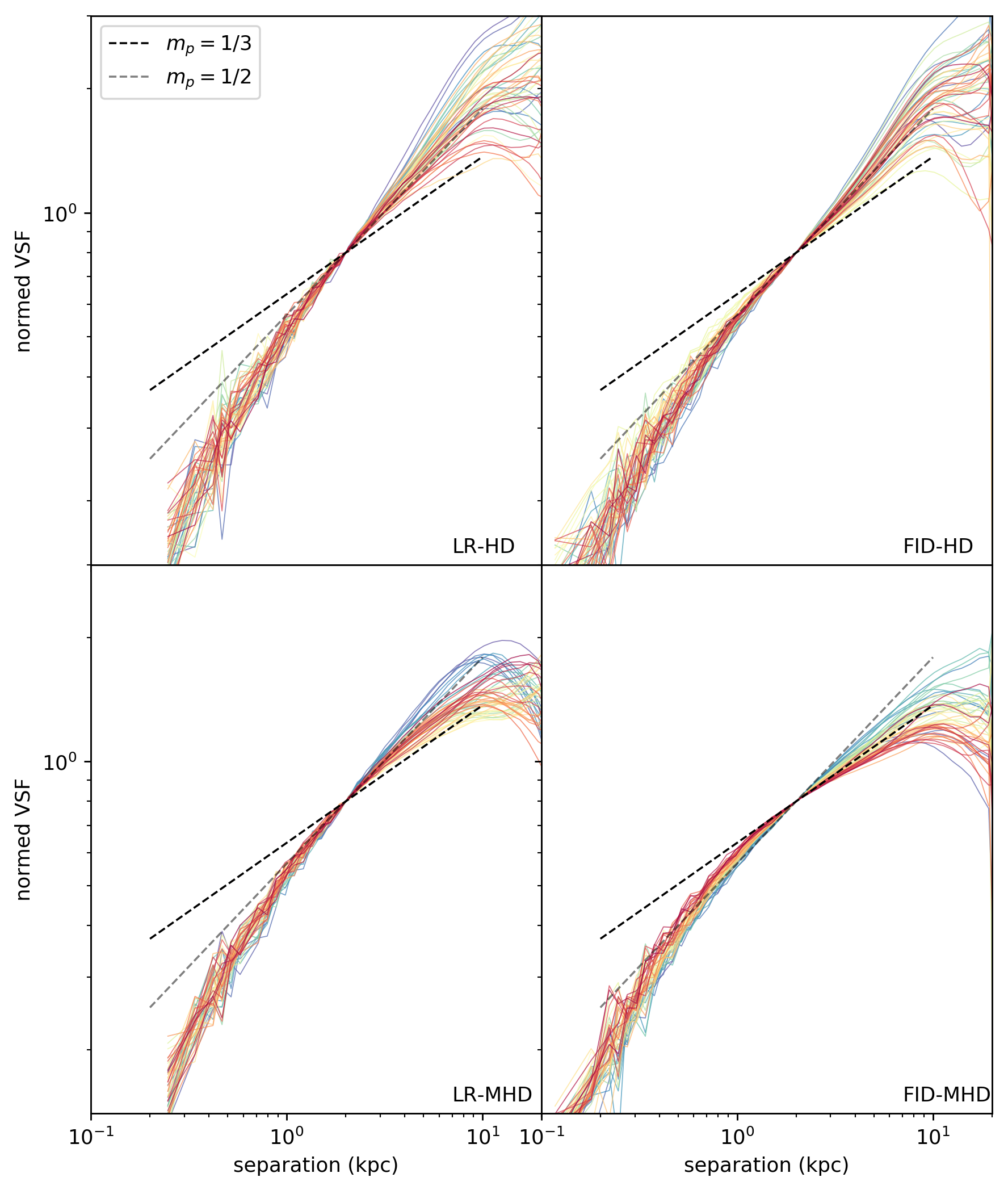}  
       \caption[]{Hot phase VSFs plotted every 10 Myr for $t=0.5 \sim 1\;{\rm Gyr}$ in four runs. Time is denoted by the color, changing from purple to red.}
     \label{fig:vsfHotActiveScatter}
  \end{center}
\end{figure}

\subsubsection{Velocity structure function of the precipitating cold gas}
Figure \ref{fig:vsfSepNames} shows the first order VSFs averaged over the AGN active phase ($t=0.5\sim1$\uu{Gyr}). 
The right four panels are the same as the left four except that they are normalized to 1 at the separation of 2\uu{kpc}. We include the normalized version in order to make it easier to compare the VSF slopes.
One of our key results is that the averaged VSF of cold filaments ($T<2\times10^{4}{\rm K}, \rho > 10^{-24}$g cm$^{-3}$) has a slope steeper than that predicted from the Kolmogorov turbulence, $m_{\rm kol}=1/3$. In the hydro cases, the slope is $m_p \approx 1/2$; and in the MHD cases, the slope is either close to $1/2$ or slightly shallower. Comparisons of blue curves in the \lrhd and \hrhd cases (top row), and of the blue curves in the \lrmhd and \hrmhd cases (bottom row), demonstrate the slopes of the cold phase VSF are numerically convergent in the inertial range ($2\sim4\uu{kpc}$ in low-resolution and $0.8\sim3\uu{kpc}$ in fiducial cases).\\
\indent  
In order to understand the nature of this scaling of the velocity with cloud separation, we inspect the acceleration terms in the Euler equation for an isolated cold filament. According to the momentum equation, the average acceleration acting on the filament can be decomposed into: gravity ($a_g$), magnetic tension ($a_t$), thermal pressure gradient ($a_p$), magnetic pressure gradient ($a_m$), and the ram pressure ($a_{\rm rp}$). The accelerations are calculated from:
\begin{equation}
        \mathbfit{a}_g = \frac{1}{M} \int \rho \mathbfit{g} {\rm d}V
\end{equation}
\begin{equation}
\mathbfit{a}_t = \frac{1}{M} \int \rho \nabla\cdot\left(\frac{\mathbfit{BB}}{4\pi}\right) {\rm d}V
\end{equation}
\begin{equation}
\mathbfit{a}_p = -\frac{1}{M} \int \nabla p_{\rm th} {\rm d}V
\end{equation}
\begin{equation}
\mathbfit{a}_m = -\frac{1}{M} \int \nabla p_{\rm mag} {\rm d}V,
\end{equation}
\begin{equation}
\mathbfit{a}_{\rm rp} = -\frac{1}{M} \rho_{\rm ICM}v_{\rm com}^2S,
\end{equation}
where the integral is over the filament; $\rho$ is the gas density; $\mathbfit{g}$ is the gravitational acceleration; $\mathbfit{B}$ is the magnetic field strength; $M=\int \rho {\rm d}V$ is the total mass of the filament; $p_{\rm th}$ is the thermal pressure; $p_{\rm mag}$ is the magnetic pressure; $\rho_{\rm ICM}$ is the density of ambient ICM; $v_{\rm com}$ is the center of mass velocity of the filament; and $S$ is the surface area of the cross section normal to $v_{\rm com}$ of the filament. We estimate $\rho_{\rm ICM}=2\times10^{-25}{\rm g\; cm}^{-3}$. As the filament falls towards the halo center, it is elongated due to the deceleration from magnetic tension force and forms a tail behind the head of the filament. The magnitudes of all the acceleration vectors are shown in Fig. \ref{fig:eulerAnalysis}. 
The solid lines are the accelerations averaged over the the head of the filament and the dashed lines are over the entire filament (note that selecting the entire filament or just its head has little effect on $a_{\rm rp}$;
filament head is defined as the location in the filament closest to the cluster center).
This analysis shows that when magnetic fields are absent, the motion of the cold filament is dominated by the gravitational force. Gravitational acceleration close to the cluster center is nearly constant. The velocity $v$ of the cold filaments subject to constant gravitational acceleration, $g$, scales with travel length $L$ as $v\propto (gL)^{1/2}$, which leads to the $m_p \approx 1/2$ slope of the VSF in the hydro runs. We verify this hypothesis by performing a simple experiment where we calculate the VSF of multiple 1D velocity-position pairs sampled from a free-fall trajectory (see Appendix \ref{app:freeFallTest} for details). When magnetic fields are included, the sub-dominant magnetic tension force increases as the filaments fall, which effectively makes the filaments gain less velocity for the same travel length compared with the hydro case. Therefore, $m_p<1/2$ for cold phase in the MHD cases.

\begin{figure}
  \begin{center}
    \leavevmode
    \includegraphics[width=\columnwidth]{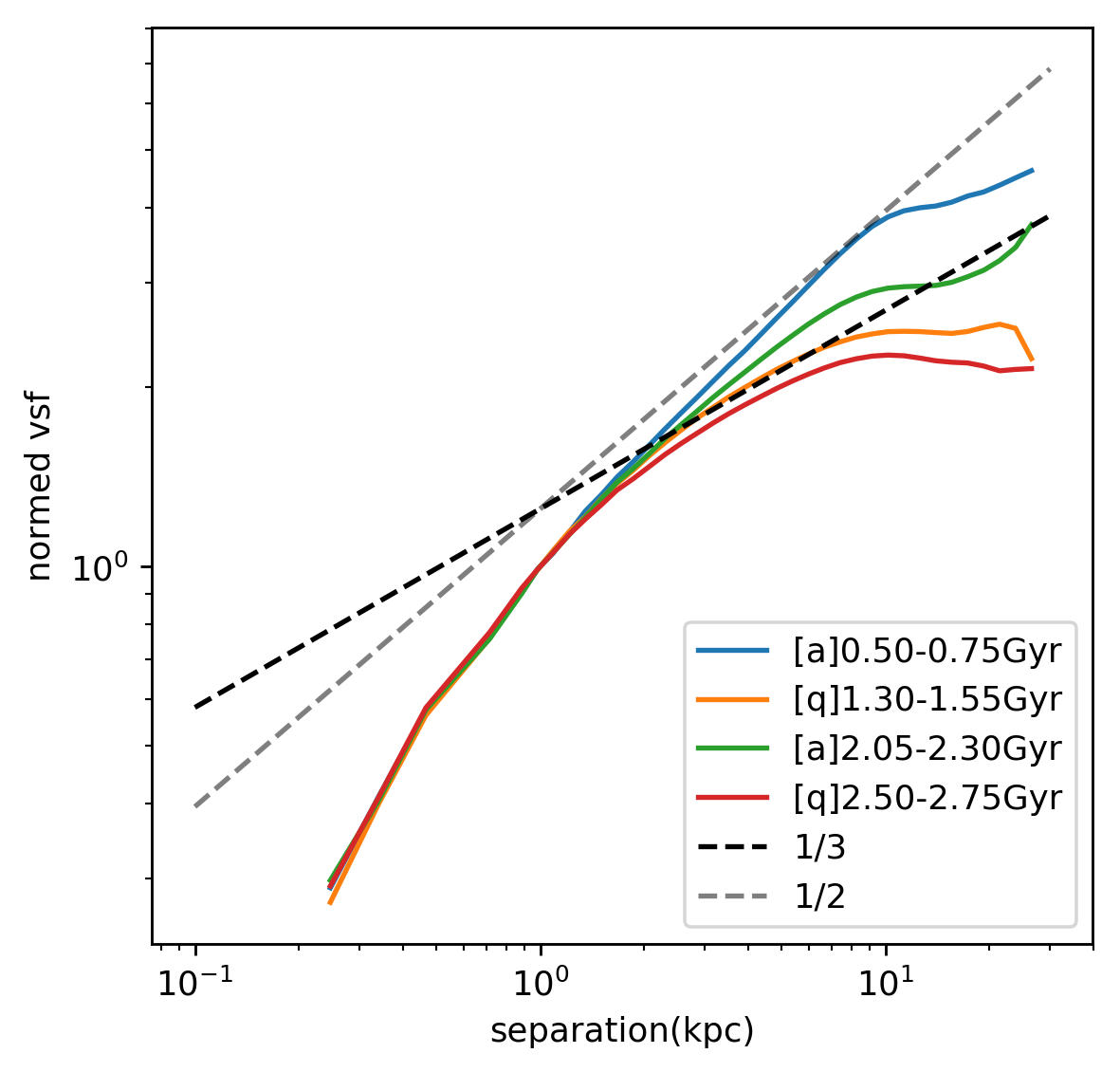}  
       \caption[]{Normalized  VSF of the hot gas averaged over quiescent and active epochs in the \lrhd case. 
       Each VSF is labelled with active (``[a]'') or quiescent (``[q]'') epoch and the corresponding time range in the legend. 
       }
     \label{fig:hvsfAtDiffEpoch}
  \end{center}
\end{figure}

\begin{figure}
  \begin{center}
    \leavevmode
    \includegraphics[width=\columnwidth]{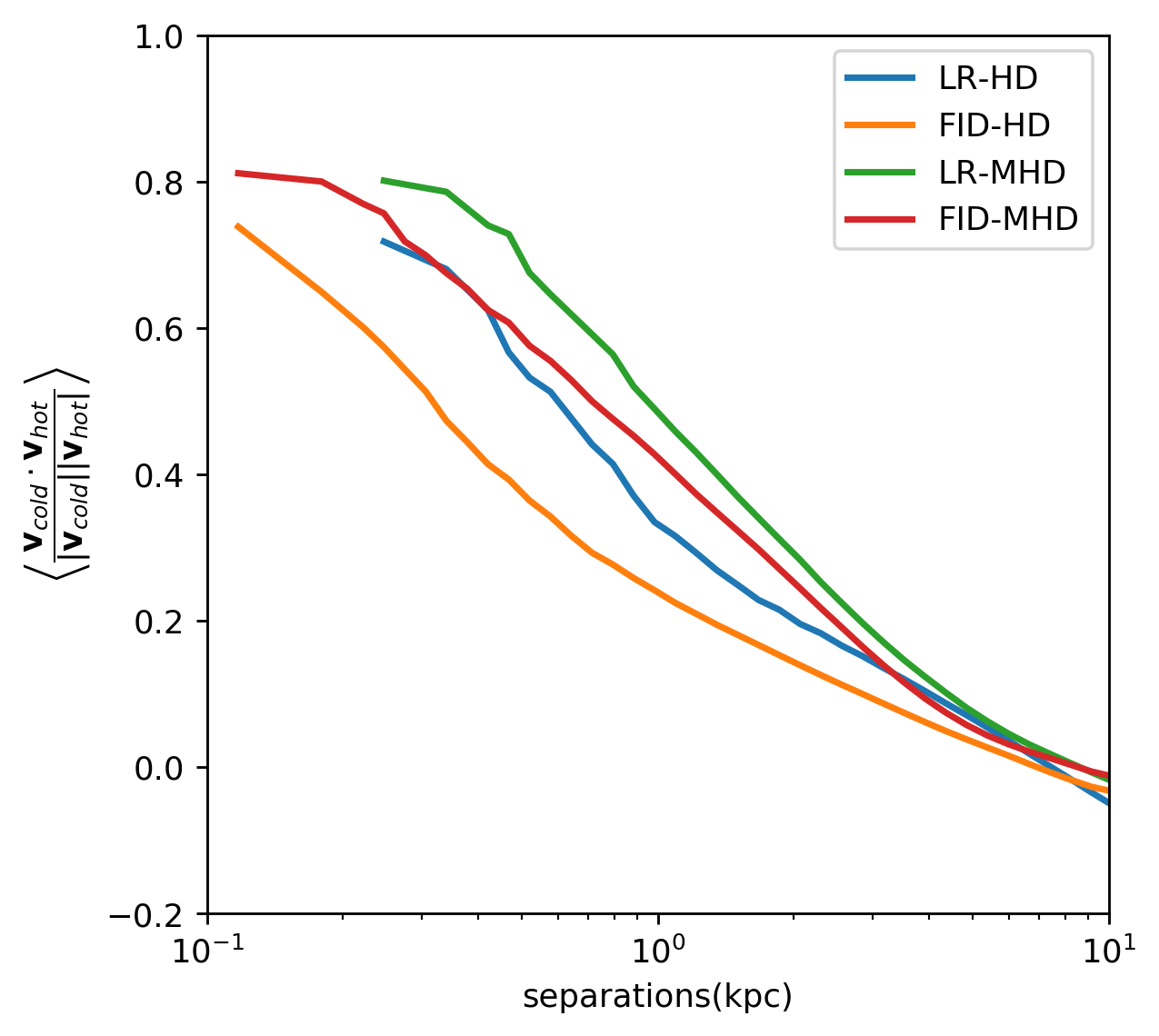}  
       \caption[]{Average spatial correlation of the cold and hot velocity normalized by the velocity magnitude.}
     \label{fig:corrColdHotVel}
  \end{center}
\end{figure}

\subsubsection{Velocity structure function of the hot ICM}
The hot phase discussed hereafter has the temperature in the range $10^{7}{\rm }{\rm K}<T<10^{8}$ K and excludes the gas component directly entrained by the bipolar jets. We define the entrained gas as that with fast outflow velocity along the jet launching axis, i.e., ${\rm sign}(z)v_z > 500\uu{km/s}$, where $v_z$ is the gas velocity along $z-$axis. \\ 
\indent
In the hydro cases, when the system is active (i.e., when the AGN jet is switched on and cold gas precipitation is present), the slope of the hot phase VSF scatters around $m_p = 1/2$. This can be seen in Fig. \ref{fig:vsfHotActiveScatter}, which shows the hot phase VSFs during the active phases (for the hydro cases, see top two panels in this figure) and the average VSF has the same slope as that of the cold phase in the inertial range (top two panels of Fig. \ref{fig:vsfSepNames}).  This result is numerically convergent and is 
consistent with the possibility that it is the cold filaments that predominantly drive turbulence in hot gas (see also Section \ref{driving sources} below). In the quiescent phase, the hot VSF flattens towards $m_{\rm kol}$ on large scales (Fig. \ref{fig:hvsfAtDiffEpoch}). We note that the same steepening/flattening trends in the active/quiescent phases are also present in the MHD cases, which we do not show in this figure in the interest of brevity.\\
\indent    
In the MHD cases, when system is in the active phase, the hot phase VSFs have slopes scattering around $m_p=1/2$ in LR-MHD case, while in the FID-MHD case, the VSFs are systematically shallower and the slopes are scattered between $m_p\lesssim 1/3$ to $m_p\approx 1/2$ (two bottom panels in Fig.  \ref{fig:vsfHotActiveScatter}). Although not numerically convergent, in both MHD cases the hot phase slopes $m_p$ are generally distributed above 1/3 predicted from Kolmogorov theory. 
In Appendix \ref{sec:appendixDrivingTurb} we verify via a simple controlled experiment that the expected slope of the hot phase VSF is indeed 1/3 and we assess at what minimum separations the numerical effects become important. 
In general, in full physics simulations, the slopes of the hot phase VSF in the MHD cases are shallower than 1/2. There appears to be tentative evidence for the flattening of the hot phase VSF in the MHD case toward the 1/3 slope expected in the Kolmogorov case. However, we note that we cannot make definite statement about the slope of the hot phase VSF due to relatively narrow inertial range and limited numerical resolution. We further discuss the driving of the hot phase turbulence in Section \ref{driving sources}.\\
\indent
As mentioned above, here we focus on three-dimensional VSF and the physics shaping the VSF slope. This allows us to {\it directly} compare simulation results to the prediction from the Kolmogorov theory. In order to make detailed comparisons to the observations of the VSF of the cold and hot phases of the ICM, one needs to perform careful projections on to the plane of the sky. As far as the hot phase VSF is concerned, \citet{ZuHone16} demonstrate that, while the projected and three-dimensional velocity power spectra have the same slope, the second order three dimensional VSF has a steeper slope than 2/3 predicted from Kolmogorov's theory. Thus, we expect that the first order projected hot phase VSF could also have a steeper slope than its Kolmogorov's counterpart. Furthermore, the exact slope may depend on factors such as, e.g., the extent of the flat core in the distribution of the X-ray emitting gas \citep{Zhuravleva12}. It will be important to understand the projection effects for X-ray emitting gas in order to compare the simulated turbulence with the observations from future high throughput high spatial resolution X-ray missions. For the cold phase, the projected slope may be shallower than the slope of the three-dimensional VSF \citep{Li20}. The relationship between the two- and three- dimensional slope may be further complicated in this case by the fact that many filaments may be seen along the same line of sight. Maps of the distribution of the cold filaments on the sky used to compute the VSF by \citet{Li20} suggest that that the covering factor may exceed unity. It is for these reasons that we defer the projection analysis to future work and in this paper we instead focus on the physics of the three dimensional VSF and comparisons to the prediction from the Kolmogorov theory.

\begin{figure}
  \begin{center}
    \leavevmode
    \includegraphics[width=\columnwidth]{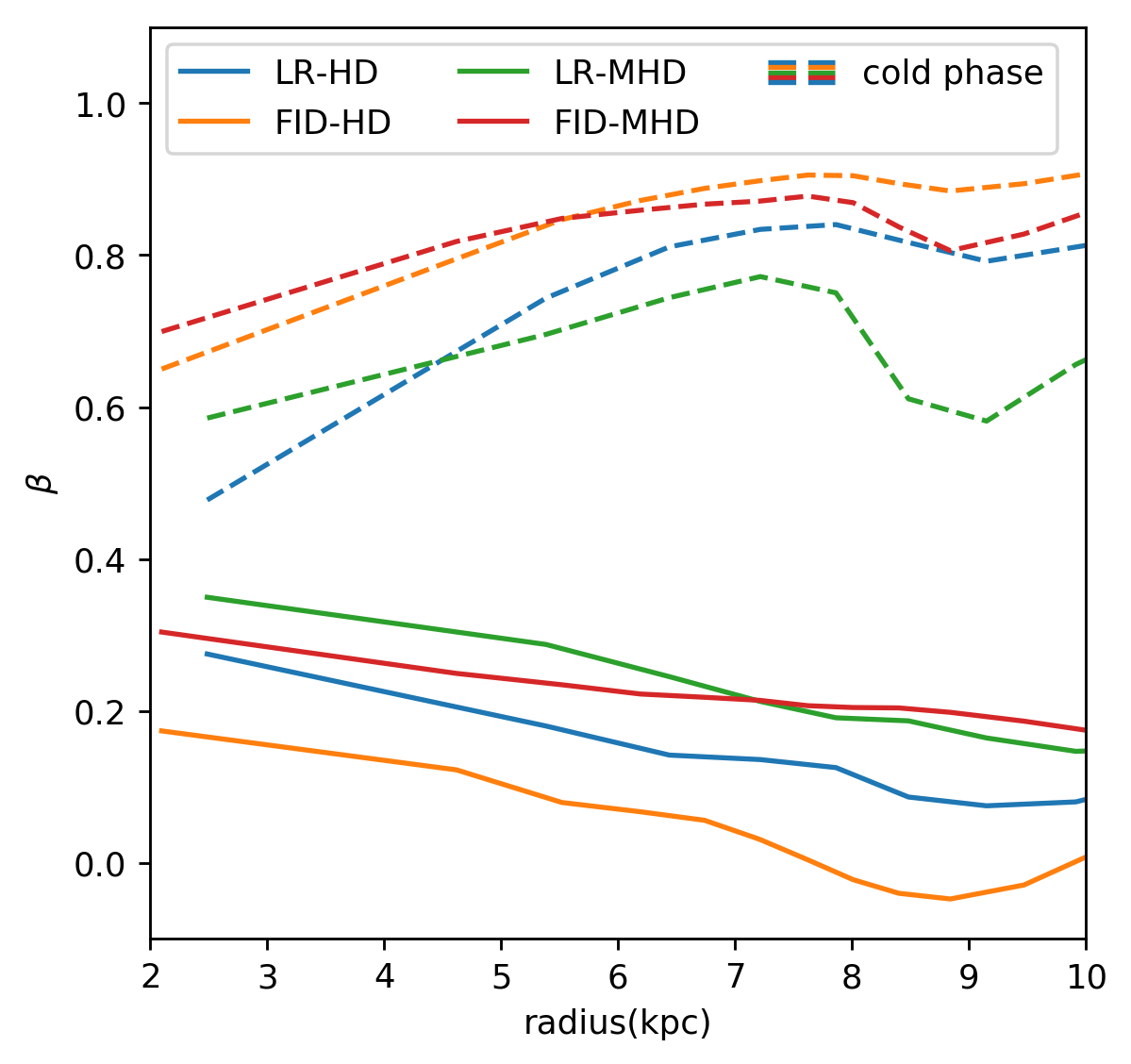}  
       \caption[]{Average radial profile of the anisotropy parameter ($\beta$) of the hot gas in all runs. $\beta$ of the cold gas is shown as dashed lines. Positive values of $\beta$ correspond to radial bias in the velocity distribution.}
     \label{fig:beta}
  \end{center}
\end{figure}

\begin{figure*}
  \begin{center}
    \leavevmode
    \includegraphics[width=0.9\textwidth]{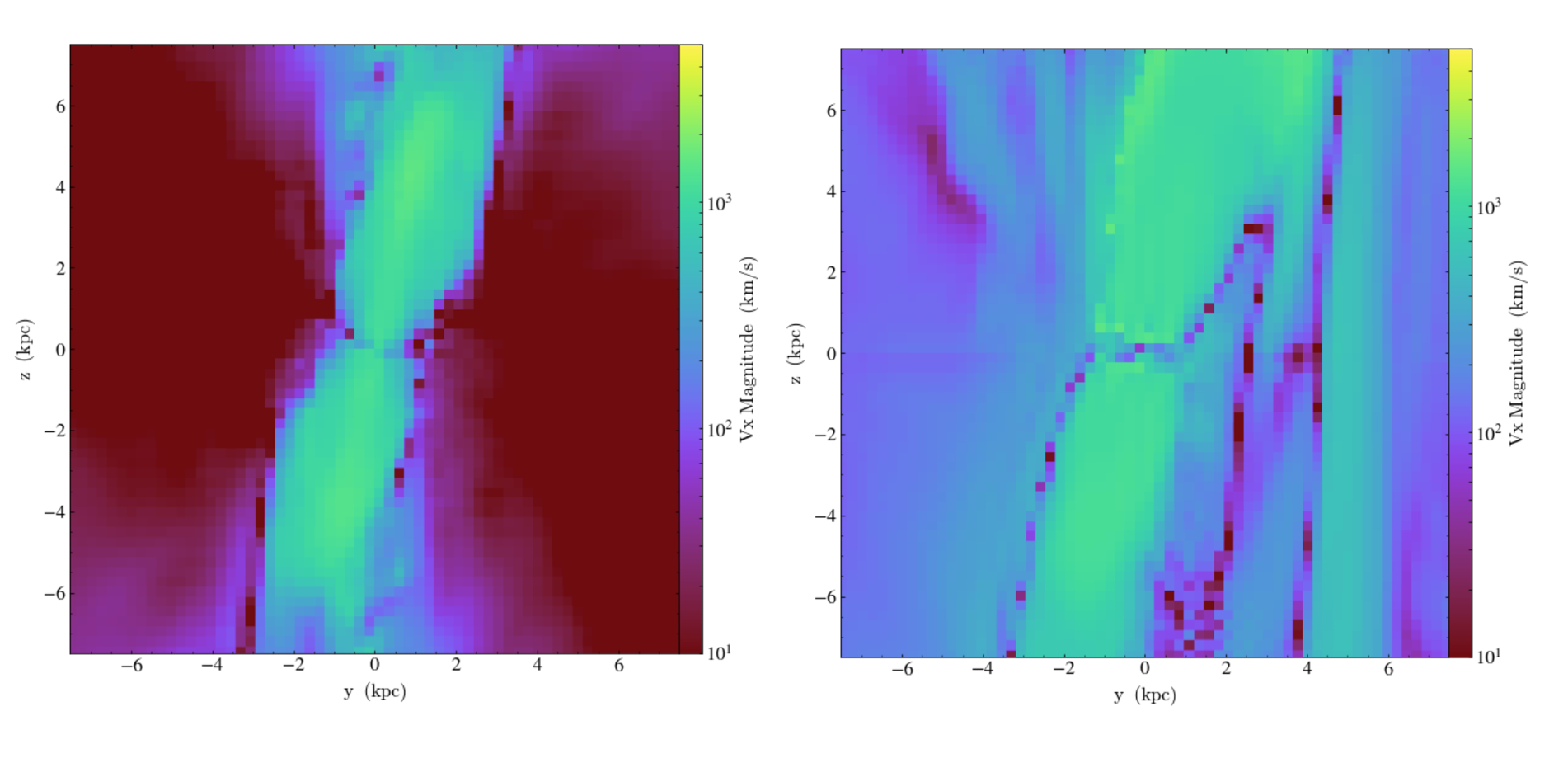}  
       \caption[]{Snapshots of the slice plot of $|v_x|$ in the test simulations with AGN manually turned-on. Left panel is for the test run with the same initial conditions as in the LR-HD run and the right panel corresponds to the same initial conditions as in the LR-MHD run.  }
     \label{fig:agnTestVxSlice}
  \end{center}
\end{figure*}

\subsection{The driving sources of turbulence in the hot ICM}\label{driving sources}

We now consider two turbulence driving mechanisms that operate in the hot phase and discuss their relative importance depending on the physics included in the simulations.

\subsubsection{Turbulence driving by cold filaments}
It is conceivable that cold filaments stir the ambient hot gas and generate turbulence. The arguments in favor of this hypothesis are the following:
\begin{itemize}
\item[(a)] The cold phase mass has a dominant contribution to the mass budget in the inner core of ICM. Specifically, as shown by the blue and red lines in Fig. \ref{fig:massDist}, the cold gas mass exceeds the hot gas mass for radii $r<10~{\rm kpc}$ in the FID cases, and we see smaller amounts of cold gas in the \hrmhd case compared to the \hrhd case. Using Chandra observations, \cite{Babyk2019} show that the ratio of molecular gas to atmospheric gas within a 10 kpc radius lies between 3\% and 50\% for central galaxies in clusters. As mentioned above, our definition of the cold gas includes the molecular gas and the gas at higher temperatures, so the cold gas portion in our simulations may be broadly consistent with the observations. Thus, in the inner region, the cold filaments have sufficient momentum and may be capable of driving the motion of the hot phase.
        
\item[(b)] Hot and cold gas velocities are spatially correlated. The spatial correlation is defined as: $f_{\rm ch}(r) = \langle{\mathbfit{v}}_{\rm hot}(\mathbfit{x})\cdot\mathbfit{{v}}_{\rm cold}(\mathbfit{x+r})\rangle_\mathbfit{x}$, where $r=|\mathbfit{r}|$. This quantity is positive and increases with decreasing spatial separation (Fig. \ref{fig:corrColdHotVel}), indicating that the velocities of the cold and hot gas parcels are more aligned when they are closer.
Thus, the motions of cold filaments and hot gas are {\it coupled}.

\item[(c)] In the hydro cases, when the system is in the active phase, the slope of hot phase VSF is $m_p\sim1/2$, which is the same as the average VSF slope of the cold gas in the inertial range.
        
\item[(d)]  The distributions of the hot and cold gas velocities are radially-biased. The anisotropy parameter is defined as: $\beta = 1-\sigma^2_{\rm tan}/2\sigma^2_{\rm rad}$, where $\sigma^2_{\rm tan} = \sigma^2_{\theta} + \sigma^2_{\phi}$. The motion of the cold filaments is highly radial due to the 
radial gravity and launching of the thermally unstable blobs by the AGN jets. The motion of the hot gas is also radial ($\beta>0$; Fig.~\ref{fig:beta}). This is so despite the fact that the hot gas entrained by the AGN jets is filtered out in the process of computing $\beta$. This implies coupling between the radially-biased cold filaments and hot phase. This conclusion is further strengthened by noting that in the process of decay of hot phase turbulence, the tangential motions should be progressively more important than the radial motions \citep[e.g.,][]{RO10} due to buoyant restoring forces. So the fact the hot gas motions are nevertheless radially-biased suggests that they may be driven by the cold gas.\\
\indent
We note that the radial bias in the velocity distribution of the hot phase is unlikely to be caused by anisotropy in the radial hot gas accretion or the ``gentle circulation'' described in \cite{YR16}. As explained below, the inflow velocities due to these two mechanisms are both much smaller than the values of the hot gas VSF (except at very small separations).\\
\indent
The average AGN cold mass accretion rates over the first Gyr of all runs are in the range $40\sim70$ $M_\odot\;$yr$^{-1}$. The estimated hot mode accretion rate is $\dot{M}\lesssim 1$ $M_\odot\;$yr$^{-1}$ as it is expected to be $\sim$2 orders of magnitude smaller than that of the cold mode \citep{GRO13}. This accretion rate corresponds to an inflow velocity $v_{\rm in, hot}\sim\dot{M}/4\pi r^2 \rho_{\rm ICM}$. For $r = 10$ ${\rm kpc}$ and $\rho_{\rm ICM}\approx10^{-25}{\rm g\; cm^{-3}}$, the inflow velocity is only $v_{\rm in, hot}\lesssim0.6$ km s$^{-1}$.\\
\indent
The ambient gas outside of the jet cone can form a reduced cooling flow \citep{YR16}. Assuming this inflow replenishes the gas mass in the central region lost to the bipolar outflow, the inflow velocity ($v_{\rm in,circ}$) can be estimated from:
\begin{equation}
    v_{\rm in,circ} = \frac{2\Omega}{4\pi - 2\Omega} \frac{\rho_{\rm jet}}{\rho_{\rm ICM}} v_{\rm jet},
\end{equation}
where $\Omega$ is the solid angle subtended by the jet cone; $\rho_{\rm jet}$ and $\rho_{\rm ICM}$ are the gas density of jet and ambient ICM respectively; and $v_{\rm jet}$ is the jet velocity. Estimating right hand side quantities at $r\approx 10\; {\rm kpc}$ to be $\Omega\approx 0.2$ (for jet cone apex angle equals to 30 degree),   $\rho_{\rm jet}/\rho_{\rm ICM}\approx0.1$, and $v_{\rm jet}\lesssim5000\;{\rm km~s}^{-1}$, the estimated inflow velocity is only $v_{\rm in,circ}\lesssim 17\;{\rm km~s}^{-1}$.

\end{itemize}

\subsubsection{Turbulence driving by AGN}
AGN jets can be an important source of turbulence in the hot phase. In order to disentangle the contributions of AGN and cold filaments to hot phase VSF, we perform four test simulations with the same initial conditions as the four full physics simulations, but excluding radiative cooling and enforcing constant AGN jet power. The constant AGN power is chosen to be consistent with the average AGN power in the production runs within the first 1 Gyr. We note that all four full physics simulations inject almost identical amount of energy during this time interval, which simplifies the interpretation of the results. \\
\indent
These test simulations demonstrate that the AGN jets effectively transfer momentum to the hot ambient gas (i.e., the hot gas outside of the jet cone) when the magnetic fields are present. Specifically, Fig.~\ref{fig:agnTestVxSlice}  shows the slice plots of $|v_x|$ for the test simulations. It is clear from this figure that the ambient gas moves much faster in the MHD case compared to the hydro case. This implies that the magnetic fields in the center of the cluster facilitate the momentum transfer from the jets to the ambient region. 

\begin{figure}
  \begin{center}
    \leavevmode
    \includegraphics[width=\columnwidth]{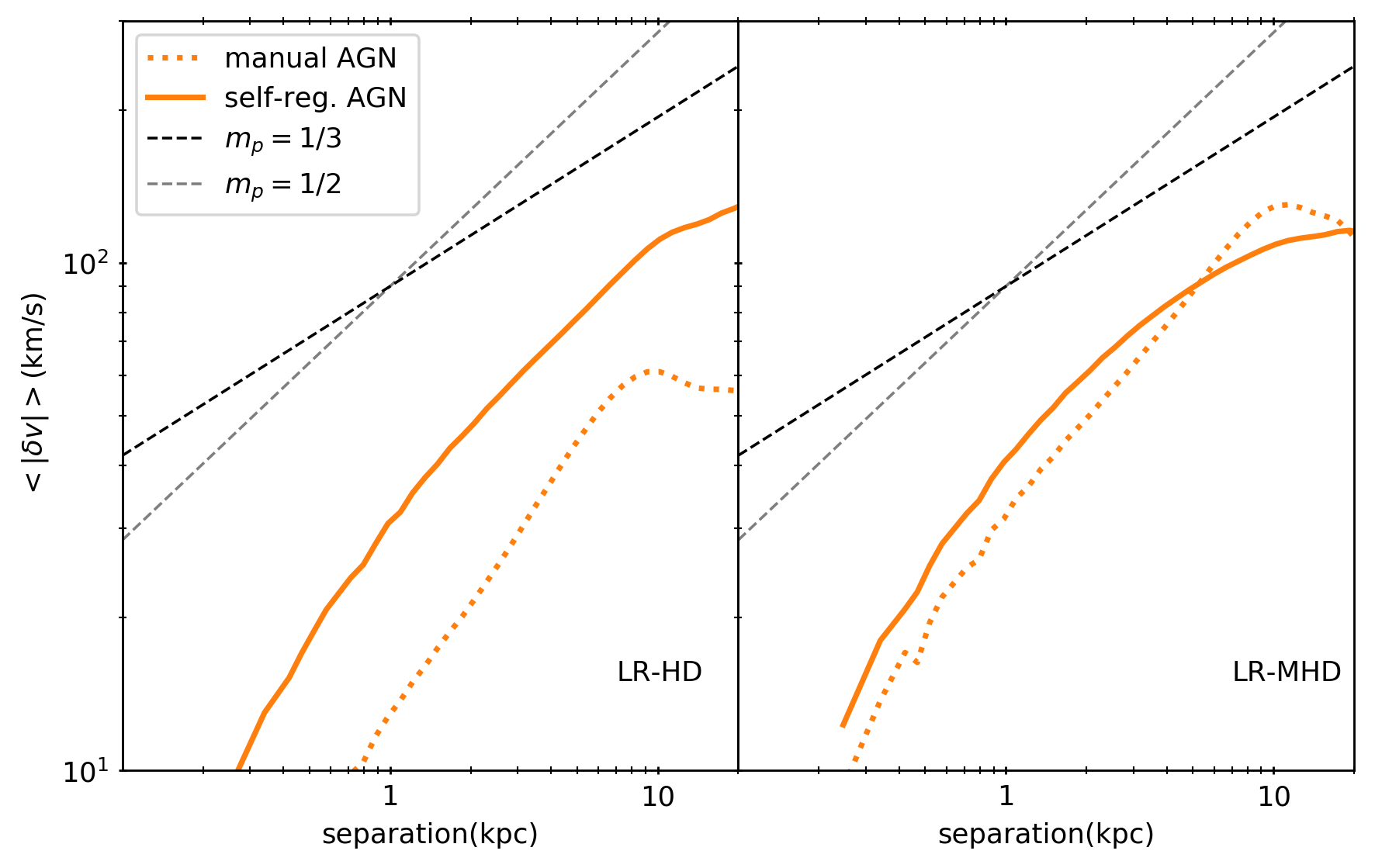}  
       \caption[]{Averaged VSF of the hot gas in self-regulated AGN simulations (solid lines) and in the corresponding test simulations with manually turned-on AGN (dot line). The VSFs are averaged over $t=0$ to 1 Gyr.}
     \label{fig:agnTestVsfCompareFullExtent}
  \end{center}
\end{figure}

Figure \ref{fig:agnTestVsfCompareFullExtent} shows the averaged VSF of hot phase gas in the production runs and those in the corresponding test runs. In this figure we consider LR-HD and LR-MHD cases as we evolved them for the longest time. However, we also note that FID-HD and FID-MHD evolved up to $\sim$0.6 Gyr were convergent when compared to their lower resolution counterparts both in terms of the normalization and slope of the VSF. In the hydro cases, the motion induced by the AGN only is subdominant (orange dotted lines in the left panel), so it is likely that the hot phase motion is mainly driven by cold filaments in this case. In the MHD cases, the VSF contribution from the AGN only (orange dotted line in the right panel) is comparable to that due to the combination of the filaments and AGN in the full physics case, indicating that the motion driver of the hot gas could be to be a mixture of the AGN jet and cold filament stirring. This is likely due to better coupling of the jets with the ambient medium facilitated by the presence of the magnetic fields. Additionally, this picture is further supported by comparing the normalization of the hot and cold VSF (left four panels of Fig. \ref{fig:vsfSepNames}). In the hydro cases, the cold phase VSF has higher velocity normalization than the hot phase VSF on all scales, so the massive cold filaments with high velocity may be sufficiently energetic to drive turbulence in the hot phase. In the MHD cases, at large scale the hot VSF has higher normalization than the cold VSF, indicating the turbulence in hot phase must be driven by other means.\\
\indent
We note that our manual jet case scenario is similar to the one considered by \citet{Hillel2020}. In their simulations, the hot phase VSF slope is 1/2, i.e., steeper than Kolmogorov, which is what we also find in our case in general. Since their simulations are too short for the local thermal instability to trigger the formation of precipitating clouds, and they neglect magnetic fields, the efficient stirring of the ICM in their case was likely facilitated by considering wide jet opening angle (half opening angle = 70 degrees) to better couple the jet to the ambient medium. In our case, direct stirring of the ICM by the jet is less efficient because the jet is narrow and the precession angle is only 15 degrees. Alternatively, jet intermittency considered by \citet{Hillel2020} could also contribute to the differences in the efficiency of the coupling of the jet energy to the ICM (perhaps even while enforcing the same average jet luminosity), though we note that more bursty AGN activity is more efficient in generating sound waves than gentler AGN activity \citep{Bambic2019}. \\

\section{Summary}
We perform 3D MHD simulations to study the properties of turbulence in the multiphase ICM affected by self-regulated AGN feedback. 
We find that, in general, the first-order velocity structure function of the multi-phase ICM is steeper than the slope predicted from Kolmogorov's theory ($m_p > 1/3$). Specifically we show that,

\begin{itemize}
    \item the turbulent motions of the cold gas are primarily driven by the gravitational acceleration. This leads to the cold phase VSF slope close to $1/2$. When magnetic fields are included, the magnetic tension decelerates the cold gas and the VSF slope is either close to 1/2 or slightly shallower. 
    
    \item Without the influence of magnetic fields, the precipitating cold filaments are likely the dominant driving source of the turbulence in the ambient hot ICM. The arguments in favor of this hypothesis are: i) in the central region of the gaseous halo, cold filaments have sufficient momentum to drive motions in the ambient hot gas; (ii) velocities of hot and cold phases are spatially correlated; (iii) in the absence of magnetic fields, the slope of hot gas VSF during AGN active phase is steeper than 1/3 and matches that of the cold phase; (iv) both hot and cold phase velocities are radially biased.
    
    \item When magnetic fields are included, turbulence in the hot phase may be driven by a combination of AGN jet stirring and filament motion. This is because the magnetic fields facilitate the AGN driving by enhanced coupling between jet-like outflow and the ambient hot gas.
    
    \item We find tentative evidence for the flattening of the hot phase VSF in the MHD case (see Fig. \ref{fig:vsfHotActiveScatter}). However, in this case, we cannot draw definite conclusions on the slope of the hot phase VSF due to narrower inertial range and lack of numerical convergence. 

\end{itemize}

\section*{Acknowledgements}
We thank Yuan Li for interesting discussions. 
Resources supporting this work were provided by the NASA High-End Computing (HEC) Program through the NASA Advanced Supercomputing (NAS) Division at Ames Research Center. MR acknowledges support from NSF Collaborative Research Grants AST-1715140 and AST-2009227, and NASA grants 80NSSC20K1541 and 80NSSC20K1583. CP acknowledges support by the European Research Council under ERC-CoG grant CRAGSMAN-646955. HYKY acknowledges support from NASA ATP (NNX17AK70G), Yushan Scholar Program of the Ministry of Education of Taiwan, and Ministry of Science and Technology of Taiwan (MOST 109-2112-M-007-037 -MY3). SPO acknowledges support from NASA grants NNX17AK58G, 80NSSC20K0539 and NSF grant AST-1911198. This research was initiated at the ``Multiscale phenomena in plasma astrophysics'' workshop at the Kavli Institute for Theoretical Physics in Santa Barbara. This research was supported in part by the National Science Foundation under Grant No. NSF PHY-1748958.



\bibliographystyle{mnras}
\bibliography{example} 




\appendix

\section{Driven turbulence test and inertial range}\label{sec:appendixDrivingTurb}

\begin{figure}
  \begin{center}
    \leavevmode
    \includegraphics[width=.9\columnwidth]{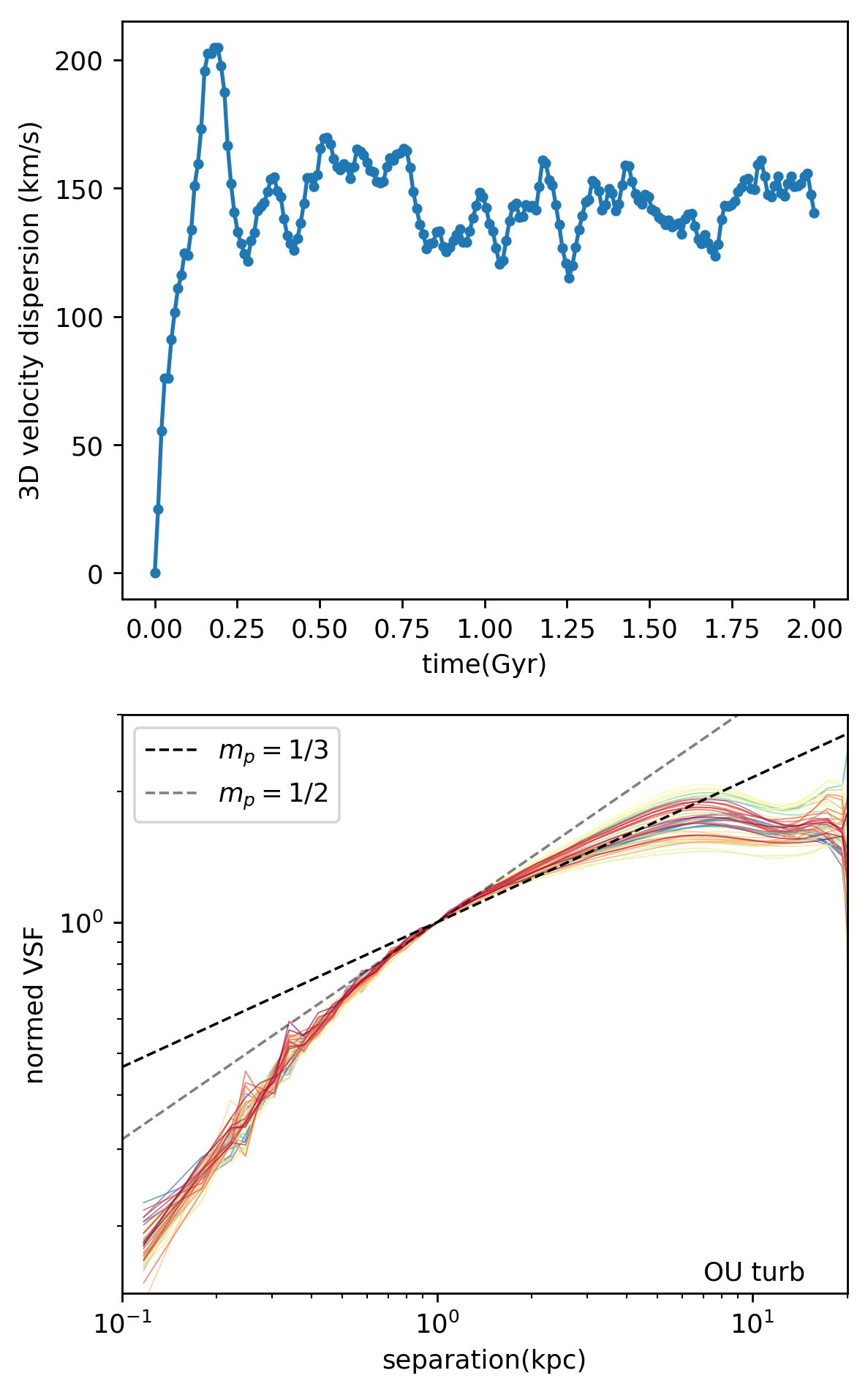}  
       \caption[]{Top panel: time evolution of the 3D velocity dispersion in the driving turbulence test (see Appendix \ref{sec:appendixDrivingTurb}). Bottom panel: VSFs plotted every 10 Myr from $t=1$ to 1.5 Gyr for the driving turbulence test. The color scheme is the same as Fig. \ref{fig:vsfHotActiveScatter}.}
     \label{fig:dataStirTest}
  \end{center}
\end{figure}

In order to verify that the expected reference value of the hot phase VSF slope is indeed 1/3, and to assess at what minimum separations numerical effects become important, we perform controlled turbulence experiment neglecting AGN feedback, magnetic fields, radiative cooling, and gravity. In this idealized test turbulence is driven via a spectral forcing scheme utilizing an Ornstein–Uhlenbeck random process. The spectral forcing scheme is set to stir the hot halo via a time-correlated, stochastic forcing with a narrow spectral range corresponding to $\sim 10$ kpc. We use the same implementation of this scheme and model parameters as described in \citet{RO10}. The computational volume in this test is uniformly refined and the resolution is the same as in the \hrhd and \hrmhd cases.

After $\sim250$ Myr, the turbulent dissipation balances the energy injection rate due to the forcing and the 3D rms velocity stabilizes at $\sim 150{\rm \;km\; s}^{-1}$ (top panel in Fig. \ref{fig:dataStirTest}). Variations are caused in part by the fact that we use finite correlation timescale for the driving forces ($\sim$100 Myr). This is reflected in the top panel in Fig. \ref{fig:dataStirTest} that shows that characteristic timescale for the fluctuations in the velocity dispersion is about 100 Myr. 

As expected, the VSF reaches a stable state, where the VSF follows the Kolmogorov prediction in the inertial range 
with the averaged slope equal to $m_{\rm kol}=1/3$. The fluctuations in the slope are again a direct consequence of the finite correlation timescale and fluctuations in the overall level of velocity dispersion. The VSF slope steepens on sub-kpc scales (bottom panel in Fig. \ref{fig:dataStirTest}). This indicates that the steepening of the VSF on sub-kpc scales is numerical in origin.

\section{VSF of idealized motion dominated by gravity}\label{app:freeFallTest}
\begin{figure}
  \begin{center}
    \leavevmode
    \includegraphics[width=\columnwidth]{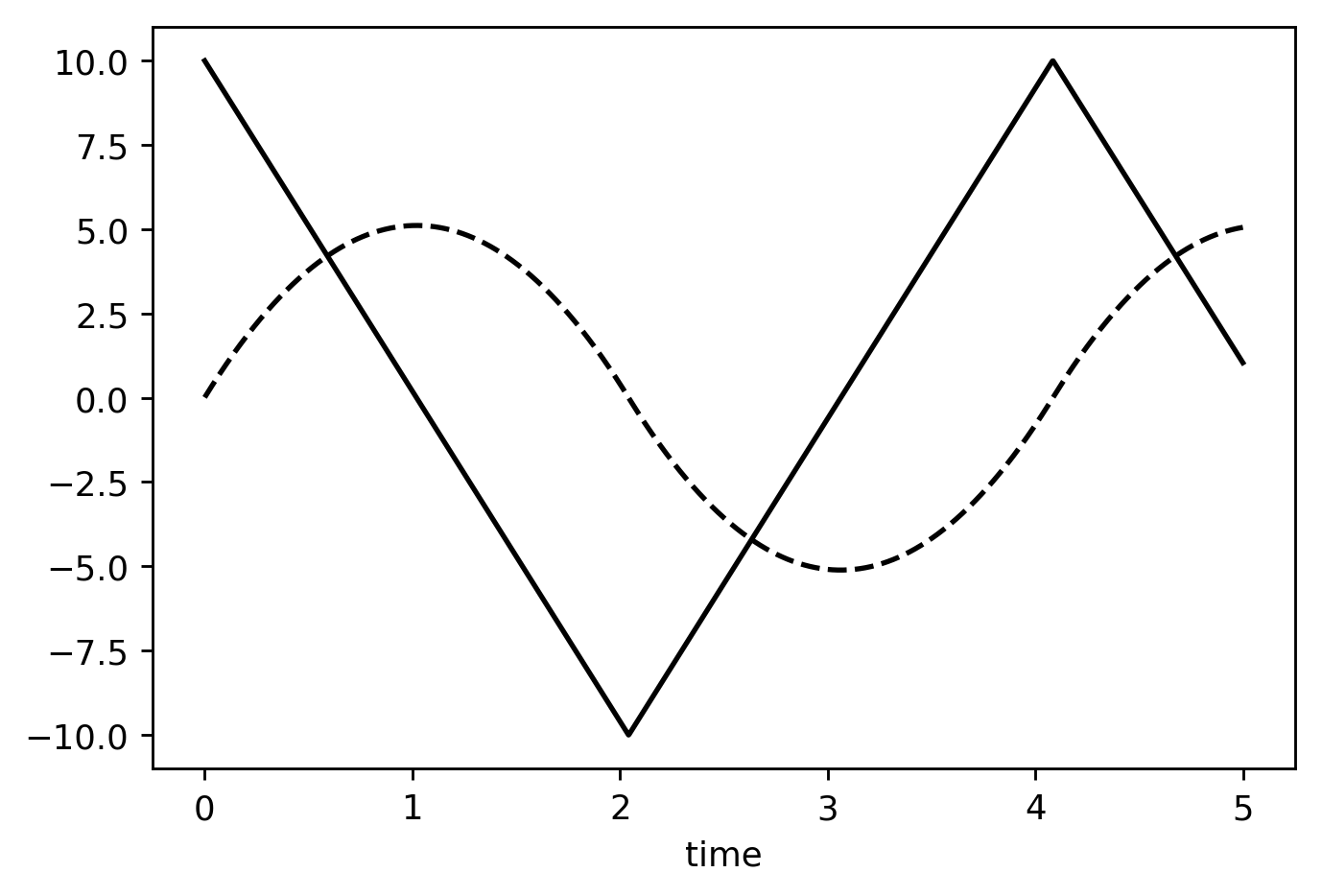}  
       \caption[]{Time evolution of velocity (solid) and position (dashed line) of a test particle in a 1D gravitational potential $g_{\rm test}(x)$ (see Appendix \ref{app:freeFallTest}). All quantities are in arbitrary units. 
       }
     \label{fig:freeFallTestVl}
  \end{center}
\end{figure}

\begin{figure}
  \begin{center}
    \leavevmode
    \includegraphics[width=\columnwidth]{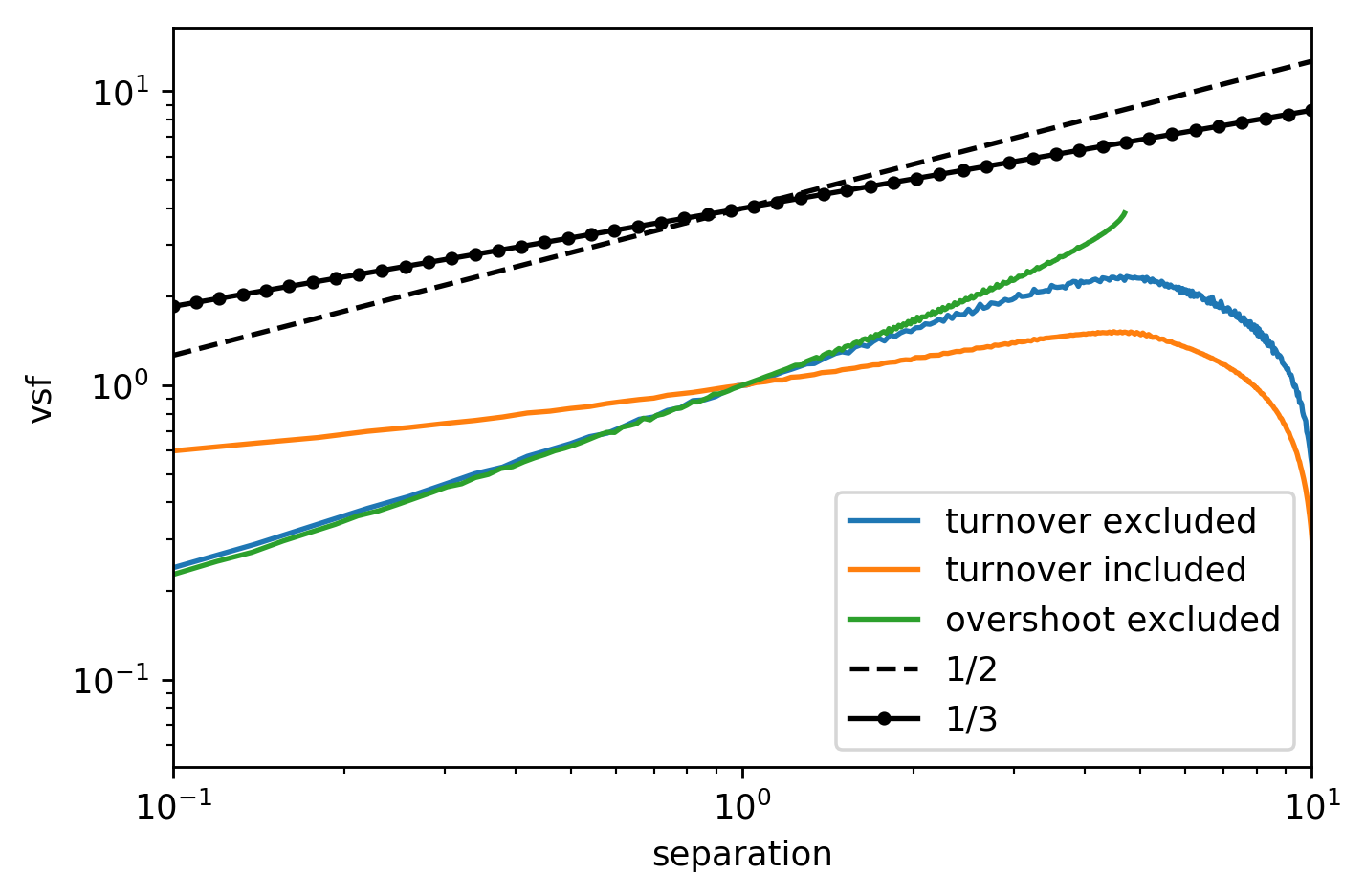}  
       \caption[]{VSF of velocities and positions sampled from the trajectory of motion dominated by gravity. The meaning of color is the same as Fig. \ref{fig:freeFallTestVl}.}   
     \label{fig:freeFallTestVsf}
  \end{center}
\end{figure}

\begin{figure}
  \begin{center}
    \leavevmode
    \includegraphics[width=\columnwidth]{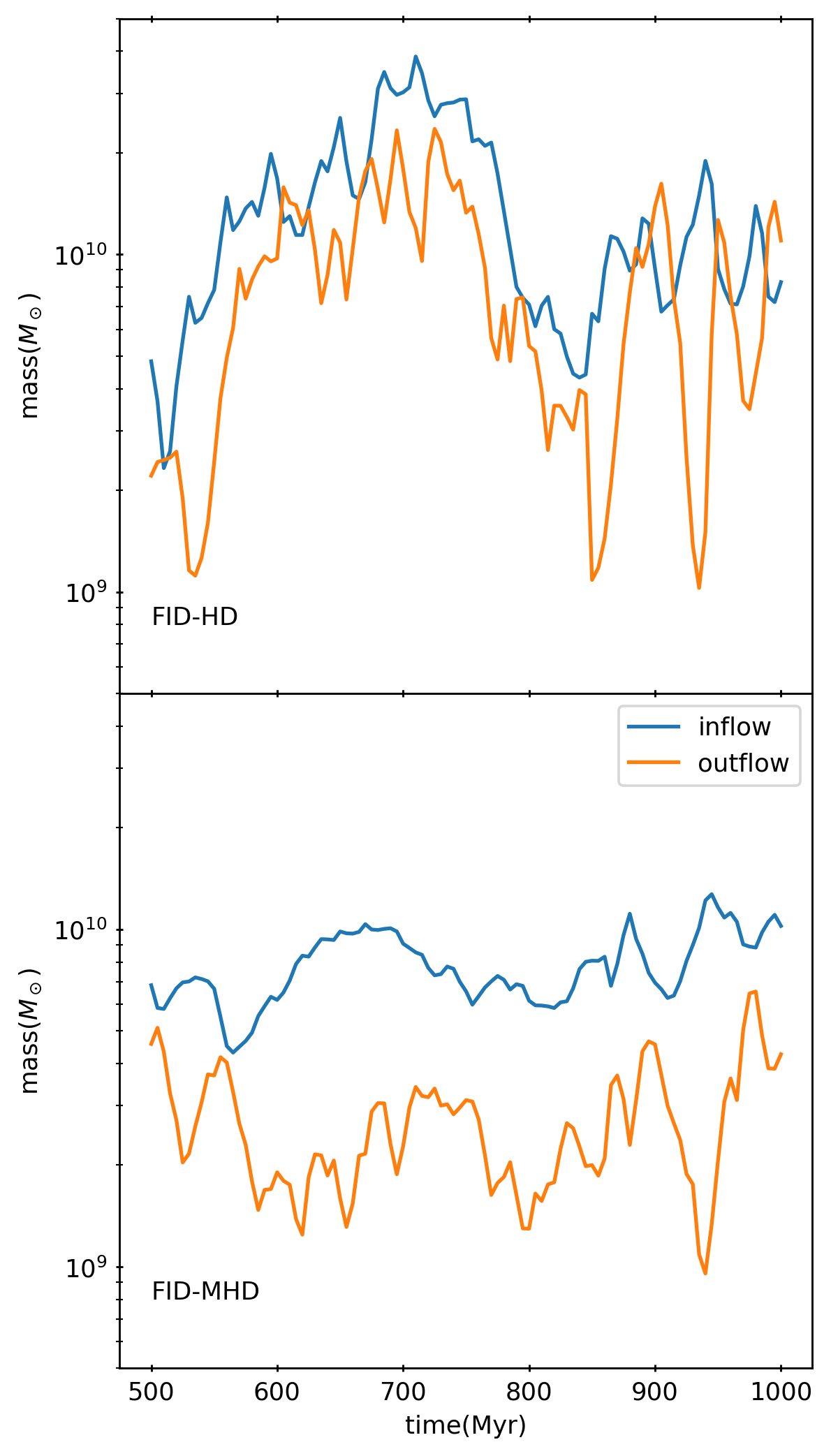}  
       \caption[]{The evolution of the inflowing (blue lines) and ouftlowing (oragne lines) cold gas mass. Top panel corresponds to the \hrhd case and the bottom one to the \hrmhd case.}   
     \label{fig:coldInOut}
  \end{center}
\end{figure}

In order to investigate the VSF slope resulting from motions dominated by gravity, we perform a simple one-dimensional calculation. We consider the motion of a test particle falling in the gravitational field $g_{\rm test}(x) = -{\rm sign}(x)g_0$, where $g_0$ is a constant (note that the gravitational acceleration in the central region of the cluster is approximately constant, which simplifies our analysis of the impact of gravity on the motions of cold gas clouds; see Fig. \ref{fig:eulerAnalysis}). The time evolution of the particle velocity and position are shown in Fig. \ref{fig:freeFallTestVl}. We sample the velocities and positions on this trajectory and calculate the VSF. In Fig. \ref{fig:freeFallTestVsf}, we show VSFs corresponding to three different sampling cases: (i) when we sample the part of the trajectory that does not include overshooting through the very center (time range from 1 to 2 in Fig. \ref{fig:freeFallTestVl}), the VSF slope is close to or slightly above 1/2 (green line in Fig. \ref{fig:freeFallTestVsf}); (ii) when we exclude the the times past the turnover where velocity changes sign (i.e., when we consider the time range from 1 to 3), the VSF slope is close to 1/2 (blue line in Fig. \ref{fig:freeFallTestVsf}); (iii) when sampling over a longer trajectory that includes the turnover (time ranging from 0 to 5), the VSF slope is close to 1/3 (orange line in Fig. \ref{fig:freeFallTestVsf}). \\
\indent
Our full physics simulation results imply that rather than getting launched by the jet from the center and raining back, most of the cold filaments form in the atmosphere and fall towards the center. In Fig. \ref{fig:coldInOut} we show the amount of cold gas that is inflowing (blue lines) and outflowing (orange lines). It is evident from this figure that the inflow dominates over outflow. The infalling cold clouds either get accreted by the black hole or collide with the preexisting cold gas clumps. Therefore, the velocity magnitude of the clouds is damped by the collisions with the preexisting cold gas as they overshoot the center of the cluster. This situation corresponds approximately to the case in between (i) and (ii) above, where the expected slope is either 1/2 or slightly above it. As our simple experiment also demonstrates that longer sampling time range results in flatter VSF slopes. As the actual magnitude of the cloud velocity is expected to be damped over time, this simple test shows that the velocities and positions sampled from a free-falling trajectory can lead to the VSF slope close to  $\sim$1/2 as seen in the full physics simulations. \\
\indent 
As mentioned in the main text, the effect of the magnetic fields is to decelerate the clouds, which may further flatten the slope below 1/2. Interestingly, the dynamical effect of the magnetic fields are also seen in Fig. \ref{fig:coldInOut}. Top and bottom panels in this figure corresponds to the \hrhd and \hrmhd cases, respectively. This figure shows that in the \hrmhd case outflow is even more subdominant compared to inflow. This is consistent with stronger damping of motions by the magnetic tension, especially as the clouds get closer to the center, that reduces the tendency of the infalling cold gas to overshoot the center.\\
\indent
Since the objective of this simple experiment is to provide a proof of concept for the idea that ballistic (or nearly ballistic) motions can account for the cold phase VSF slopes, we choose not do pursue a more detailed analysis including additional free parameters such as the cloud velocity damping time or the distributions of heights from which the clouds are released, etc.

\bsp	
\label{lastpage}
\end{document}